\documentclass[epj]{svjour}
\usepackage{graphics}
\usepackage{epsfig}
\newcommand{\vev}[1]{\langle#1\rangle}
\newcommand{\Fs}[1]{F_{#1,\mathrm{short}}}
\newcommand{\Fl}[1]{F_{#1,\mathrm{long}}}
\newcommand{\Tr}{\mathrm{Tr}}

\begin{document}
\title{Heavy quark potential and quarkonia dissociation rates}
\author{D. Blaschke\inst{1,2}
\thanks{\emph{Present address: GSI mbH, 
D-64291 Darmstadt, Germany}}%
\and O. Kaczmarek\inst{1}
\and E. Laermann\inst{1}
\and V. Yudichev\inst{2}
}                     
\offprints{D. Blaschke}          
\institute{$^1$Fakult\"at f\"ur Physik, Universit\"at Bielefeld,
D-33615 Bielefeld, Germany\\
$^2$Bogoliubov Laboratory of Theoretical Physics, JINR Dubna, 141980 Dubna,
Russia}
\date{Received: date / Revised version: date}
%
\abstract{
Quenched lattice data for the $Q\bar Q$ interaction (in terms of heavy quark 
free energies) in the color singlet channel at finite temperatures are fitted 
and used within the nonrelativistic
Schr\"odinger equation formalism to obtain binding energies and scattering 
phase shifts for the lowest eigenstates in the charmonium and bottomonium 
systems in a hot gluon plasma.
The partial dissociation rate due to the Bhanot-Peskin process is calculated 
using different assumptions for the gluon distribution function, including 
free massless gluons, massive gluons, and massive damped gluons. 
It is demonstrated that a temperature dependent 
gluon mass has an essential influence on the heavy quarkonia dissociation,
but that this process alone is insufficient to
describe the heavy quarkonia dissociation rates.
} 
\maketitle
\section{\label{sec:intro}Introduction}

Heavy quarkonia have been suggested as hard probes of the quark gluon plasma
\cite{Matsui:1986dk} since the modification of static 
interactions at finite temperature 
eventually implies a dissolution of heavy quarkonia bound states into the 
continuum of scattering states (Mott effect). This effect results in a 
suppression of heavy quarkonia production in heavy-ion collisions as an 
observable signal. 
Since the Mott temperatures for J/$\psi$, $\Upsilon$ and $\Upsilon'$ 
as obtained by solving the Schr\"odinger equation for a screened 
Cornell-type potential lie well above the critical temperature $T_c$ for 
deconfinement  \cite{Karsch:1987pv}, it has soon been realized that
a kinetic theory is necessary for the description of heavy quarkonia 
dissociation \cite{Ropke:1988bx}, see \cite{Burau:2000pn,Grandchamp:2002wp} 
for recent formulations. 
Solutions of the Schr\"odinger equation provide the basis for the 
evaluation of  cross sections and rates for the Bhanot-Peskin process 
\cite{Bhanot:1979vb} of heavy quarkonia dissociation by 
gluon impact \cite{Blaschke:2004dv}.
In this contribution we present a new fit to the singlet free energies from 
quenched lattice QCD simulations \cite{Kaczmarek:2002mc,Kaczmarek:2004gv} 
and show that binding energies and cross 
sections deviate from those obtained for Debye potential fits 
\cite{Digal:2001ue,Wong:2004zr,Arleo:2004ge}. 

\section{Heavy quark potential}
The main source of our knowledge of the quark-antiquark
static interaction at high temperature originates from  calculations of
the free energy for a quark-antiquark system
from the Polyakov-loop correlator from lattice QCD.
The color averaged  correlator receives contributions from
the color singlet and color octet channels
\begin{equation}
\vev{\Tr[L(0)]\Tr[L^\dagger(r)]}=
\frac19\exp\left(-\frac{F_1(r)}{T}\right)+\frac89
\exp\left(-\frac{F_8(r)}{T}\right).
\end{equation}
The color-singlet part $F_1$ is extracted from the equation
\begin{equation}
\vev{\Tr[L(0)L^\dagger(r)]}=
\exp\left(-\frac{F_1(r)}{T}\right)
\end{equation}
and the color octet part is found by subtraction.
The free energy of a quark-antiquark system thus obtained is interpreted
hereafter as the effective static interaction potential
for a quark-antiquark pair surrounded by gluons. 

\subsection{Zero temperature}

\begin{figure}[tb]
\hspace{-0.02\textwidth}
\includegraphics[width=0.46\textwidth,angle=-90]{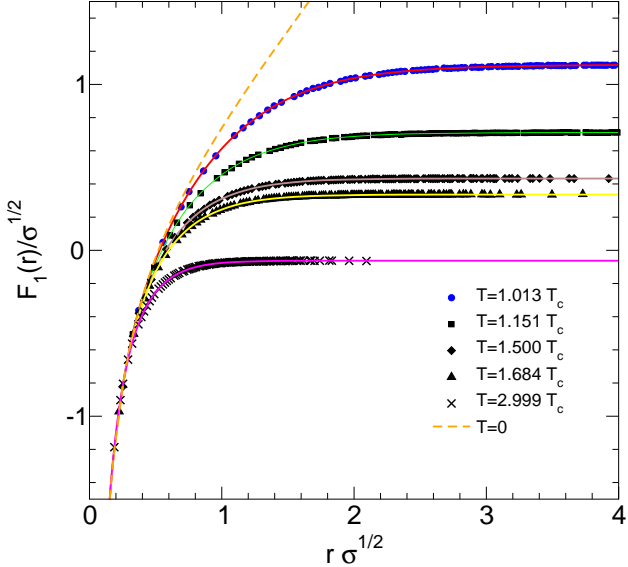}
\caption{Lattice data for the singlet free energy \cite{Kaczmarek:2004gv} at 
different temperatures, see legend. Solid lines are obtained with the fit
on the basis of the Dixit formula, dashed line is the $T=0$ potential.}
\label{fig:1}       
\end{figure}

As there exist no color octet mesons in the vacuum, we will use the color
singlet free energies as a potential in the Schr\"odinger equation 
description of the heavy quarkonia spectrum. Two additional parameters are 
to be defined in fitting the spectra: the heavy quark masses
and a constant shift of the whole potential.
The behavior of the quark-antiquark interaction in the color singlet channel 
was investigated in Ref. \cite{Necco:2001xg} within the combined lattice and
perturbative QCD approach. We fit these data points implementing a $\chi^2$ 
minimization to the Ansatz
\begin{equation}
\label{F1}
F_1(r)=\left\{
\begin{array}{ll}
\Fs{1}(r), & r<r_0,\\
\Fl{1}(r), & r\geq r_0,
\end{array}\right.
\end{equation}
where $\Fs{1}(r)$ describes the interaction at short distances
whereas $\Fl{1}(r)$ is responsible for the long-dis\-tan\-ce forces.
Both expressions are matched at $r=r_0$ that is defined below.
This point, as we shall see, lies in the domain of perturbative QCD.
We use the combined linear and Coulomb potential to describe the
long-distance interaction and the Coulomb interaction with the
$r$-dependent coupling constant $\alpha(r)$ for  short distances
\begin{eqnarray}
\Fl{1}(r)&=&\sigma r-\frac{\pi}{12r},\\
\Fs{1}(r)&=&-\frac43\frac{\alpha(r)}{r},\\
\alpha(r)&=& \frac{16\pi}{33}\left(
    \frac{1}{\ln\left(r^2/c^2\right)} -\frac{r^2}{r^2-c^2}\right).
\end{eqnarray}
The formula for $\alpha(r)$ is obtained by solving the one-loop
renormalization group equation for the running coupling constant in QCD
followed by the pole subtraction \cite{Shirkov:2002gw}.
The constant $c\sqrt{\sigma}\approx 1.816$
and the point $r_0\sqrt{\sigma}\approx 0.031$ are determined 
in units of the string tension $\sqrt{\sigma}= 0.42$~GeV from the
condition that the potential is a smooth function at $r=r_0$.
The result is given by the dashed line in Fig.~\ref{fig:1}.

\subsection{Singlet free energy at high temperature }
For the singlet free energy  of a static quark-antiquark system at high 
temperature we use the Ansatz
\begin{equation}
F_1(r,T)=\left\{
\begin{array}{ll}
\Fs{1}(r), & r<r_0,\\
\Fl{1}(r,T), & r\geq r_0,
\end{array}\right.
\end{equation}
where for the short-range interaction governed by pQCD the zero-temperature
form of the previous subsection is assumed.

The long-distance interaction $\Fl{1}(r,T)$ requires theoretical
assumptions about its shape. Instead of the frequently used screened
Coulomb potential, we follow  Dixit \cite{Dixit:1989vq} and
assume the  potential behavior at large $r$ as
$V\sim \exp(-\mu r^2)/\sqrt{r}$.
Above $T_c$, the expression for the potential without the
Coulomb interaction was deduced in
\cite{Dixit:1989vq}
\begin{eqnarray}
\Fl{1}^{(0)}(r,T>T_c)&=&
 -\frac{q^2}{2^{3/4}\Gamma(3/4)}\sqrt{\frac{r}{\mu}}
 K_{1/4}\left[\left(\mu r\right)^2\right]\nonumber\\
&&+ q^2\frac{\Gamma(1/4)}{2^{3/2}\Gamma(3/4)\mu},
\end{eqnarray}
and contains two parameters: $q$ and $\mu$.
A similar parameterization was used by Digal et al. \cite{Digal}.
We add to this term a Coulomb attraction at short distances and obtain
\begin{equation}
\Fl{1}(r,T>T_c)=\Fl{1}^{(0)}(r,T>T_c)+\Fs{1}e^{-(\mu r)^2}.
\end{equation}
The parameters $q$ and $\mu$ are determined from a fit to lattice data 
\cite{Kaczmarek:2004gv}, the results are shown in Fig. \ref{fig:1}.

\section{Quarkonia in a hot gluon plasma}
\subsection{Quarkonia at zero temperature}
The masses of quarkonia in the vacuum are defined as
\begin{equation}
M=2m_Q+E+v_0
\end{equation}
where $m_Q$ is the quark mass, and the energy $E$ is an eigenvalue 
of the Schr\"odinger equation
\begin{equation}
\label{se}
[-\nabla^2/m_Q+V(r)]\psi(r)=E\psi(r)~,
\end{equation}
where the potential $V(r)$ is identified with the zero temperature free
energy $F_1(r)$ of Eq. (\ref{F1}) up to an unknown constant $v_0$.
Substituting the wave function 
$\psi_{n\ell{}m}(r,\theta,\phi)=r^{-1}R_{n\ell}(r)Y_{\ell{}m}(\theta,\phi)$
into (\ref{se}), one obtains an equation for $R_{n\ell}(r)$.
At large $r$, the potential is linear, and the solution
of 
this equation behaves as the Airy function
$R_{n\ell}(r)\sim \mathrm{Ai}(\kappa r -\xi)$,
where $\kappa^3=m_Q\sigma$ and $\xi=m_Q E/\kappa^2$.

The masses of 1S and 4S states are used as input.
For charmonium we obtain $m_c=1.45~\mbox{GeV}$ and the constant
$v_0= -302$~MeV. For the bottomonium we have $m_b=4.785$~GeV with the same 
$v_0$. Once $m_c$, $m_b$, and $v_0$ are fixed, the remaining quarkonia 
spectrum is described \cite{Blaschke+05}.

\subsection{Quarkonia at finite temperature}

\begin{figure}[tb]
\includegraphics[width=0.32\textwidth,height=0.6\textwidth,angle=-90]{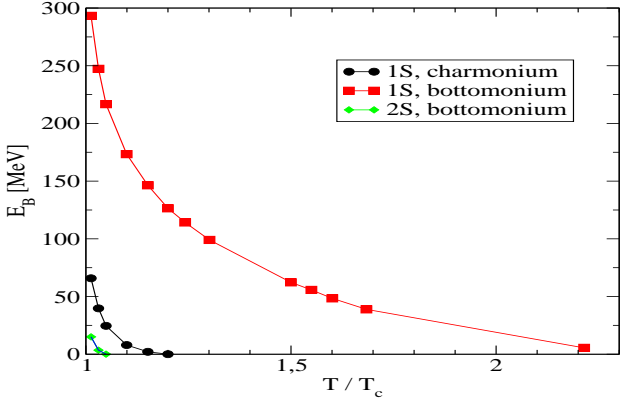}
\caption{Binding energies of heavy quarkonia states from solution of the
Schr\"odinger equation for the $T$-dependent effective potential of Fig. 
\ref{fig:1}.}
\label{fig:2}       
\end{figure}

The Schr\"odinger equation for a bound state in the QGP has the form
\begin{eqnarray}
[-\nabla^2+m_Q{V}_{\rm eff}(r,T)]\psi(r,T)&=&m_Q E_B(T)\psi(r,T),
\end{eqnarray}
where $E_B(T)>0$ is the temperature dependent binding energy.
The medium effects on the $Q\bar Q$ system are modeled using the singlet free 
energies as an effective potential ${V}_{\rm eff}(r,T)=F_1(r,T)-F_{\infty}(T)$
with the continuum threshold $F_\infty(T)=\lim_{r\to\infty} F_1(r,T)$.
The temperature dependent mass of a quarkonium bound state is defined as
\begin{equation}
M(T)=2m_Q-E_B(T)+v_0+F_{\infty}(T)~.
\end{equation}
The solutions for
the binding energy both for  charmonium and bottomonium are
shown in Fig.~\ref{fig:2}.

\begin{figure}[tb]
\includegraphics[width=0.46\textwidth,angle=-90]{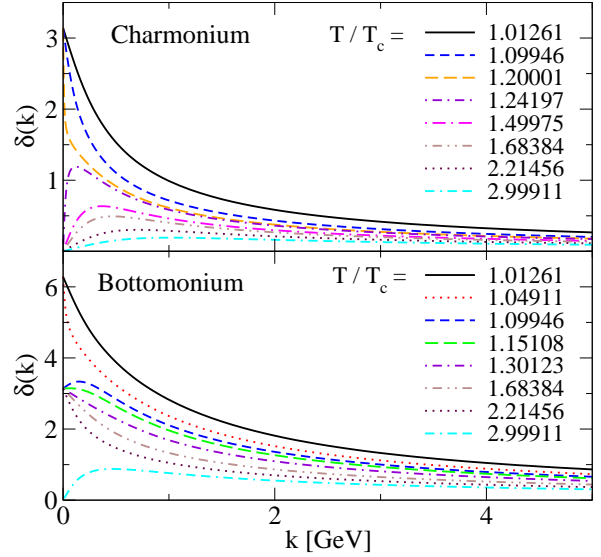}
\caption{Scattering phase shifts for heavy charmonia and bottomonia 
states from solution
of the Schr\"odinger equation for the $T$-dependent potential of
Fig. \ref{fig:1}.}
\label{fig:3}       
\end{figure}

The wave function of an unbound quark-antiquark system can be
calculated via the S-wave phase shift function $\delta_S(r)$  
by solving the equation \cite{Calogero:67}
\begin{equation}
\frac{d\delta_S(k,r,T)}{dr}=-\frac{m_Q V_{\rm eff}(r,T)}{k}
\left[\sin(kr+\delta_S(k,r,T))\right].
\end{equation}
The phase shift is defined as $\delta_S(k,T)\equiv \delta_S(k,\infty,T)$
and results are shown in Fig. \ref{fig:3} for charmonia and bottomonia states
at different temperatures. In accordance with the Levinson theorem, the 
scattering phase shift at threshold changes by $\pi$ once a bound state merges 
the continuum at its Mott temperature; $T^{\rm Mott}/T_c=1.05,~1.20,~2.25$
for $\Upsilon'$, J/$\psi$, $\Upsilon$, respectively, see Figs. 2 and 3.

\section{Dissociation of quarkonia by gluon impact}
\begin{figure}[tb]
\includegraphics[width=0.46\textwidth,angle=-90]{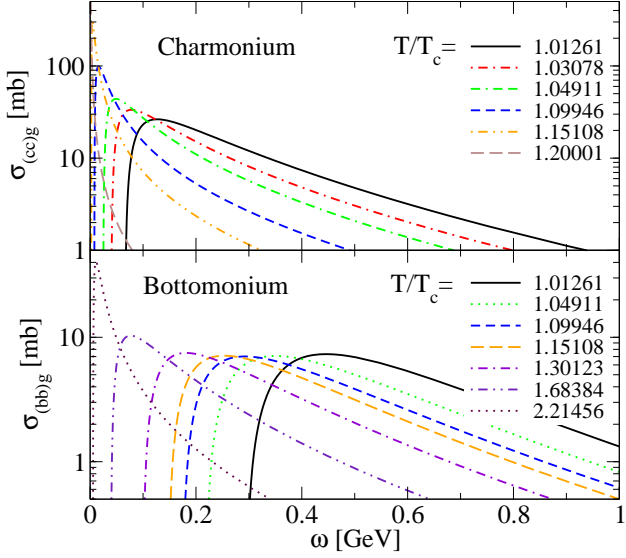}
\caption{Cross sections for J/$\psi$ and $\Upsilon$ dissociation
by the Bhanot-Peskin process.}
\label{fig:5}       
\end{figure}

We calculate the cross section for the quarkonium dissociation
after a gluon impact similarly to the calculation of
the deuteron decay via the photon absorption \cite{Wong:2004zr,Blatt:52}
\begin{equation}\label{dis}
\sigma_{(Q\bar Q)g}(\omega)=
\frac{4\pi\alpha_{gQ}}{3}\frac{(k^2+k_0^2)}{k}\left(
\int\limits_{0}^{\infty} u_{1P}(r)u_{1S}(r)\, r\; dr\right)^2,
\end{equation}
\begin{equation}\label{norm}
\int\limits_{0}^{\infty}|u_{1S}(r)|^2 dr =1,
~ \alpha_{gQ}={\alpha_s}/{6},~k_0^2=m_Q E_B(T),
\end{equation}
where we used $R_{n\ell}(r) =u_{n\ell}(r)e^{-k_0r}$.
For the 1P state, one can use the wave function of a free
quark-antiquark system:
\begin{equation}
u_{1P}(r)=\frac{\sin kr}{kr}-\cos kr, \quad k^2=m_Q (\omega-E_B(T)).
\end{equation}
For the constant $\alpha_s$ in (\ref{norm}), we take an
average over the low energy region below 1 GeV, which
gives $\alpha_s\approx 0.48$.
As a result, we obtain the cross sections shown in
Fig.~\ref{fig:5}. Their peak values correspond to the geometrical 
ones ($\pi R^2(T)$) with the T-dependent radius $R(T)$
of the quarkonia wave function \cite{Blaschke+05}.

\begin{figure}[tb]
\epsfig{figure=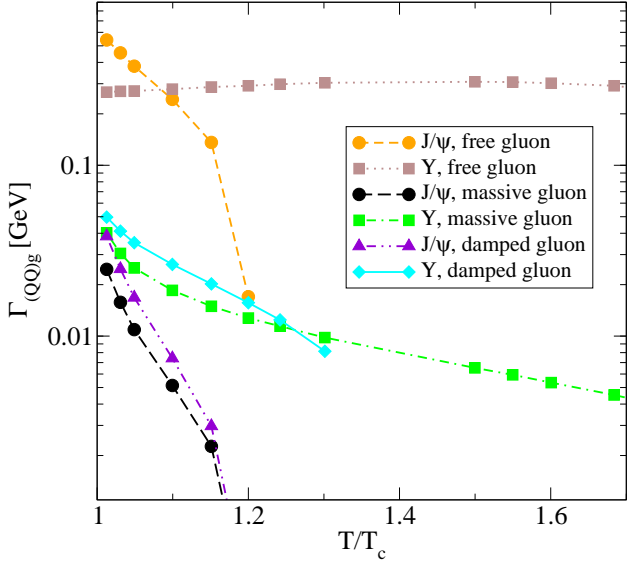,width=0.46\textwidth,angle=-90}
\caption{Dissociation rates of heavy quarkonia 
in a hot gluon plasma
obtained with massive damped, massive undamped and massless (free) gluon 
distributions in Eq. (\ref{Gamma}), respectively.}
\label{fig:7}       
\end{figure}

Now we can estimate the dissociation rate of the charmonium and
bottomonium by gluon impact according to
\begin{equation}
\label{Gamma}
\Gamma_{(\bar QQ)g}(T)=\int\limits_{0}^{\infty}\!
ds A(s)\int\limits_{{s}}^{\infty}\!
\frac{d\omega^2}{4\pi^2}\sqrt{\omega^2-s}\,
\sigma_{(Q\bar Q)g}(\omega) n_g(\omega),
\end{equation}
where  $A(s)$ is the normalized gluon spectral function
\begin{equation}
A(s)=\frac{1}{\pi}\frac{\sqrt{s}\gamma}{(s-m_g^2)^2+s\gamma^2},\qquad
\int\limits_{0}^{\infty} ds A(s)=1,
\end{equation}
and the thermal gluon distribution function is given by
$
n_g(\omega)=2(N_c^2-1)[\exp(\omega/T) - 1]^{-1}.
$
The temperature dependent gluon mass $m_g$ and damping width $\gamma$ 
are taken from a recent fit to lattice QCD data for the entropy in 
pure gauge  \cite{Peshier:2004bv},
\begin{eqnarray}
m_g^2=2\pi\bar\alpha \,T^2,~~
\gamma= 3\bar\alpha\, T \ln({2.67}/{\bar\alpha}),
\end{eqnarray}
where $\bar\alpha=12\pi/[33\ln(3.7(T-T_s)/T_c)^2]$ with $T_s=0.46$ GeV. 
The results are shown in Fig. \ref{fig:7} and compared with the cases 
when the damping width and also the gluon mass are neglected.

\section{Conclusions}
We have used a new fit to recent quenched lattice data for the $Q\bar Q$ 
singlet free energies at finite temperatures  to obtain binding energies 
and scattering phase shifts for the lowest eigenstates in the charmonium 
and bottomonium systems within the Schr\"odinger equation formalism.
In contrast to results on the basis of a Debye 
potential fit, we obtain much smaller finite temperature quarkonia binding 
energies, entailing large dissociation cross sections for the Bhanot-Peskin 
process.
The corresponding dissociation rates have been evaluated using different 
assumptions for the gluon distribution function, including free massless 
gluons, massive gluons, and massive damped gluons. 
We have demonstrated that that a temperature dependent 
gluon mass has an essential influence on the heavy quarkonia dissociation.
However, the Bhanot-Peskin process alone is insufficient to
describe the quarkonium dissociation process 
\cite{Grandchamp:2002wp,Blaschke:2004dv}.
On the basis of the spectrum and wave functions obtained here we will study 
next the $Q\bar Q$ spectral functions above the Mott temperature and compare 
the results with corresponding lattice studies \cite{Karsch:2005ex}.
\subsection*{Acknowledgements}
{V.Yu.} received support from the Heisenberg-Landau programme, the Dynasty 
Foundation and RFBR grant No. 05-02-16699.
This work has been supported in part by the Virtual Institute of the Helmholtz 
Association under grant No. VH-VI-041.

\end{document}